\documentclass[a4paper,11pt]{article}

\usepackage{pos,physics,float}
\usepackage{subfigure,graphicx,epsfig}
\usepackage{natbib,amsmath}

\newcommand{\mycomment}[1]{}

\title{Eigenspectra of Minimally Doubled Fermions}

\author*[a]{Abhijeet Kishore}
\author[b]{Subhasish Basak}
\author[a]{Dipankar Chakrabarti}

\affiliation[a]{Department of Physics, Indian Institute of Technology
Kanpur, Kanpur-208016, India}
\affiliation[b]{School of Physical Sciences, NISER Bhubaneswar and
Homi Bhabha National Institute, India}

\emailAdd{akishore@iitk.ac.in}
\emailAdd{sbasak@niser.ac.in}
\emailAdd{dipankar@iitk.ac.in}

\abstract{
In this work, we explored the eigenspectra of minimally doubled fermions,
in both Karsten-Wilczek and Borici-Creutz realizations. We
generated 4-dim $SU(3)$ gauge fields with a definite topological charge
and calculated the chiralities of the eigenmodes for KW and BC fermions.
We used the spectral flow of the eigenvalues for this purpose and demonstrated the Index theorem.}

\FullConference{The 41st International Symposium on Lattice Field
Theory (LATTICE2024)\\
 28 July - 3 August 2024\\
Liverpool, UK\\}

\begin{document}
\maketitle

\section{Introduction}
Minimally doubled fermions (MDF) are proposed as the local actions
with chiral symmetry on the lattice. In the presence of a background
gauge field with an integer-valued topological charge, they should
satisfy the Atiyah-Singer index theorem. Two popular MDF formulations,
namely Karsten-Wilczek (KW)~\cite{Karsten:1981gd,Wilczek:1987kw} and
Borici-Creutz (BC)~\cite{Creutz:2007af,Borici:2007kz} fermions in
2-dim are shown to satisfy the index theorem~\cite{Chakrabarti:2009sa,
Creutz:2010bm,Durr:2022mnz,Pernici:1994yj,Tiburzi:2010bm}. To make
a case for MDFs in full QCD simulations, it is important to investigate
the same in 4-dim. In this work, we investigate the index theorem
for both KW and BC fermions in 4-dim with $SU(3)$ gauge field, having
a specific topological charge in the background by employing the spectral
flow of the eigenvalues of the corresponding Hamiltonian. We observe
that $\gamma_5$ is not the appropriate operator to identify the
chirality of the zero modes. The modified chirality operators with
the flavored mass term have been defined for both KW and BC fermions,
which are found to reproduce correct chiral states of the
doublers or taste as it is often called.

\section{KW and BC fermions}\label{sec:fermion}
According to Nielsen-Ninomiya theorem, chiral fermion on lattice comes
with an extra number of fermion species or doublers. The minimum number of
doubler fermions on the lattice one can have is two. The two most popular
variants of minimally doubled fermions (MDF) are Karsten-Wilczek and
Borici-Creutz fermions~\cite{Karsten:1981gd,Wilczek:1987kw,
Creutz:2007af,Borici:2007kz}. Both actions add terms to the naive
fermion action that anticommutes with $\gamma_5$, and hence, the changed
actions remain chiral, each with two chiral zero modes. In KW fermions,
a Wilson-like term with only spatial derivatives multiplied by
$i\gamma_4$ is added to the free Dirac operator. In 4-dim discrete
spacetime lattice, the KW action in the presence of a background gauge field is
\begin{equation}
\begin{split}
S_{\text{KW}} = \sum_{\substack{x}}\Big[ \frac{1}{2} \sum_{\substack{
\mu=1}}^{4} \bar{\psi}(x) \gamma_{\mu} \Big\{ U_{\mu}(x) \psi(x+
\hat{\mu}) - U^{\dagger}_{\mu}(x-\hat{\mu}) \psi(x-\hat{\mu}) \Big\}\\
- \frac{i}{2} \sum_{\substack{\mu=1}}^{3} \bar{\psi}(x) \gamma_{4}
\Big\{ U_{\mu}(x) \psi(x+\hat{\mu}) - 2\psi(x)  + U^{\dagger}_{\mu}(
x-\hat{\mu}) \psi(x-\hat{\mu})\Big\} + m\bar{\psi}(x) \psi(x) \Big]
\end{split}    \label{kw_action}
\end{equation}
The appearance of zero modes in two different places in the Brillouin zone
becomes apparent in momentum space. In free massless theory ($U=1$), the KW
operator in momentum space is,
\begin{eqnarray}
D_{\text{KW}}(p) &=& \sum_{\substack{\mu=1}}^{4} i
\gamma_{\mu} \, \sin p_{\mu} + 2i \gamma_4 \sum_{\substack{\mu=1}}^{3}
[\sin (p_{\mu}/2)]^2 \label{kw_mom_dirac}
\end{eqnarray}
Inverting the momentum space Dirac KW operator,
the zeros show up at $(0,0,0,0)$ and $(0,0,0,\pi)$ which corresponds to
the two doublers. Having a preferred time direction, it breaks the
hypercubic symmetry of the naive lattice action. Alternatively, the
MDF action proposed by Borici and Creutz on 4-dim orthogonal lattice
is written as,
\begin{equation}
\begin{split}
S_{\text{BC}} = \sum_{\substack{x}}\Big[ \frac{1}{2} \sum_{\substack{
\mu=1}}^{4} \bar{\psi}(x) \gamma_{\mu} \Big\{ U_{\mu}(x) \psi(x+\hat{\mu})
- U^{\dagger}_{\mu}(x-\hat{\mu}) \psi(x-\hat{\mu}) \Big\}\\
+ \frac{i}{2} \sum_{\substack{\mu=1}}^{4} \bar{\psi}(x) \gamma'_{\mu}
\Big\{ U_{\mu}(x) \psi(x+\hat{\mu}) - 2\psi(x)  + U^{\dagger}_{\mu}
(x-\hat{\mu}) \psi(x-\hat{\mu})\Big\} + m\bar{\psi}(x) \psi(x) \Big]
\end{split} \label{bc_action}
\end{equation}
where $\gamma'_{\mu}=\Gamma - \gamma_{\mu}$, $\;\Gamma=(1/2)
\sum_{\substack{\mu}} \gamma_{\mu}$.
The two zeros of BC operator in momentum space are at $(0,0,0,0)$
and $(\frac{\pi}{2},\frac{\pi}{2},\frac{\pi}{2},\frac{\pi}{2})$ as is
evident from the poles of the inverse of Dirac BC operator in the
momentum space,
\begin{eqnarray}
D_{\text{BC}}(p) &=& \sum_{\substack{\mu}} i
\gamma_{\mu} \, \sin p_{\mu} - 2i \sum_{\substack{\mu}} \gamma'_{\mu} \,
[\sin (p_{\mu}/2)]^2  \label{bc_mom_dirac}
\end{eqnarray}
The BC operator satisfies the $\gamma_5$-hermiticity, and like KW, is
diagonal in momentum space.

Some exploratory studies of minimally doubled fermions, both KW and BC,
have been carried out, like mixed action spectroscopy~\cite{Basak:2017oup,
Godzieba:2024uki,Weber:2013tfa,Weber:2015hib,Weber:2016dgo}, eigenspectra~\cite{Durr:2020yqa}, taste structures~\cite{Ammer:2024yro,Durr:2024ttb,Weber:2023kth}, chiral symmetry breaking~\cite{Osmanaj:2018pqb}, and index theorem~\cite{Creutz:2010bm,Durr:2022mnz}.
Some studies of formal aspects include renormalization properties~\cite{Capitani:2010nn,Vig:2024umj,Capitani:2013zta,Capitani:2013fda,Capitani:2009ty,Capitani:2009yn,Capitani:2010ht,Osmanaj:2022mzs,Kimura:2011ik}, phase structure~\cite{Misumi:2012uu,Kimura:2012df,Misumi:2012ky}, anomaly structure~\cite{Tiburzi:2010bm},
and construction of chiral perturbation theory~\cite{Shukre:2024tkw}. 
However, for MDF to become a serious contender for QCD simulation, we
feel that more analytical and numerical studies are needed, particularly in the topological
sector. In the present work, we have studied the spectral flow
of the eigenvalues with respect to the flavored mass~\cite{Creutz:2010bm,
Durr:2022mnz} and index theorem of KW and BC action in 4-dim with a
definite topological charge for background $SU(3)$ gauge fields.

\section{Spectral flow and the index theorem}\label{sec:specflow}
Atiyah-Singer Index Theorem relates the difference of numbers of
left and right-handed (\textit{i.e.,} of opposite chiralities) zero
eigenmodes of massless Dirac operator $D$ to topological charge
$Q_{\text{top}}$,
\begin{eqnarray}
\text{index}(D) &=& n_{+} - n_{-} = (-1)^{d/2}Q_{\text{top}} \label{qtop} \\
\text{index}(D_{\text{mdf}}) &=& 2 \times (-1)^{d/2}Q_{\text{top}}
\label{qtopmdf}
\end{eqnarray}
The $Q_{\text{top}}$ is a property of the gauge fields. In eqn. (\ref{qtopmdf}), an extra factor of 2 arises from two fermion species~\cite{Creutz:2010bm} of MDF. We obtained the
$\text{index}(D)$ by spectral flow~\cite{Edwards:1998sh,Creutz:2010bm},
where the flow of eigenvalues of the Dirac operator is considered as a
function of both the bare and flavored mass parameters. The $\text{
index}(D)$ is calculated by counting the net number of times the eigenvalues
change sign in the flow near $m=0$, counted with sign $\pm$ depending on the slope of crossings, and equals $n_+ - n_-$. For the computation of the eigenspectrum, we use
a combination of Kalkreuter-Simma algorithm~\cite{Kalkreuter:1995mm} for
eigenvectors and LAPACK for eigenvalues. We implemented the
algorithm by suitably modifying appropriate subroutines in the publicly
available MILC code~\cite{MILC}. We computed eigenspectra for $D_{\text{mdf}} + M$ and its hermitian counterpart, where mdf stands for either KW or BC and $M$ for either bare mass $(m)$ or \textit{flavored mass}.
The hermitian version of MDF with flavored mass is defined as
\cite{Creutz:2010bm,Durr:2022mnz},
\begin{eqnarray}
H_{\text{KW}}(m) &=& \gamma_5 (D_{\text{KW}} + m [C_{\text{sym}}\otimes1])
\label{hkw5} \\
H_{\text{BC}}(m) &=& \gamma_5 (D_{\text{BC}} + m [(2C_{\text{sym}} - 1)
\otimes1]) \label{hbc5}
\end{eqnarray}
where 
\begin{eqnarray} 
C_{\text{sym}} &=& \frac{1}{4!} \sum_{\substack{\text{perm}}} C_1
C_2 C_3 C_4,\\
\text{and}~~ C_{\mu}(x,y) \,\psi(y) &=& \frac{1}{2} \big[
U_{\mu}(x) \delta_{x+\hat{\mu},y} + U^{\dagger}_{\mu}(x-\hat{\mu})
\delta_{x-\hat{\mu},y} \big]\psi(y).
\end{eqnarray}
We calculated numerically the eigenvalue flows
of $H_{\text{KW}}(m)$ and $H_{\text{BC}}(m)$ on a background $SU(3)$
gauge field configuration with a fixed topological charge $Q_{\text{top}}$.
For this, we followed the proposal given in reference~\cite{Gattringer:1997c,
Smit:1986fn}. The simplest gauge
field with a non-zero $Q_{\text{top}}$ is when $F_{\mu\nu}(x)$ is
a constant. On a 4-dim torus with $x_{\mu} \in [0,L_{\mu}]$, the link
variables, $U_{\mu}(x)$ are
\begin{eqnarray}
U_1(x) &=& exp(-i\omega_1 a x_2 \tau_j), \;\;\text{and} \;\; U_2(x) =
\begin{cases}
1, & \text{for} \, x_2 = 0,a,...,(N-2)a \\ 
exp(i\omega_1 L x_1 \tau_j), & \text{for} \, x_2=(N-1)a
\end{cases} \label{u1u2} \\
U_3(x) &=& exp(-i\omega_2 a x_4 \tau_j), \;\; \text{and} \;\; U_4(x) =
\begin{cases}
1, & \text{for} \, x_4 = 0,a,...,(N-2)a \\
exp(i\omega_2 L x_3 \tau_j), & \text{for} \, x_4=(N-1)a
\end{cases} \label{u3u4} \\
Q_{\text{top}} &=& 2n_1 n_2 \;\;\; \text{where}, \;L=Na, \;\;\omega_i a^2
= 2\pi n_i/L^2 \label{uqtop}
\end{eqnarray}
The topological charge $Q_{\text{top}}$ for a smooth $SU(3)$ gauge field is,
\begin{equation}
Q_{\text{top}} = -\frac{1}{4\pi^2} \sum_{\substack{x}} \, \text{Tr}
\big[F_{12}(x) F_{34}(x) - F_{13}(x)F_{24}(x) + F_{23}(x)F_{14}(x) \big]
\label{qtop_def}
\end{equation}
where $F_{\mu\nu}(x)$ is the field strength tensor. 
The generated field corresponding to different $Q_{\text{top}}$
are considered smooth in the sense they are not elements of Markov chains. 
A smooth configuration does not always give well-separated sign flips
for the eigenvalues, and hence we roughen the $U_\mu(x)$ by keeping
$Q_{\text{top}}$ approximately invariant,
\begin{equation}
U_{\mu}(x)_{(\delta)} = exp(i\sum_{\substack{j}} \theta_{\mu}^{(j)}
\tau_j) \implies U_{\mu}(x)_{\text{rough}} = U_{\mu}(x)_{(\delta)}
U_{\mu}(x)_{old} \label{roughU}
\end{equation}
where $U_{\mu}(x)_{old}$ are smooth links (\ref{u1u2}, \ref{u3u4})
and $U_{\mu}(x)_{(\delta)}$ are $SU(3)$ elements in the vicinity of
identity. $\tau_j$ are Gell-Mann matrices and $\theta_{\mu}^{(j)}$
$j=1,2,\ldots,8$ are small random numbers uniformly distributed in
$(-\delta\pi, \delta\pi)$. In reference~\cite{Smit:1986fn}, the
roughening is done to mimic Monte Carlo configurations to increase
statistics.
\begin{figure}[h]
\makebox[\textwidth]{
\includegraphics[width=0.45\textwidth]{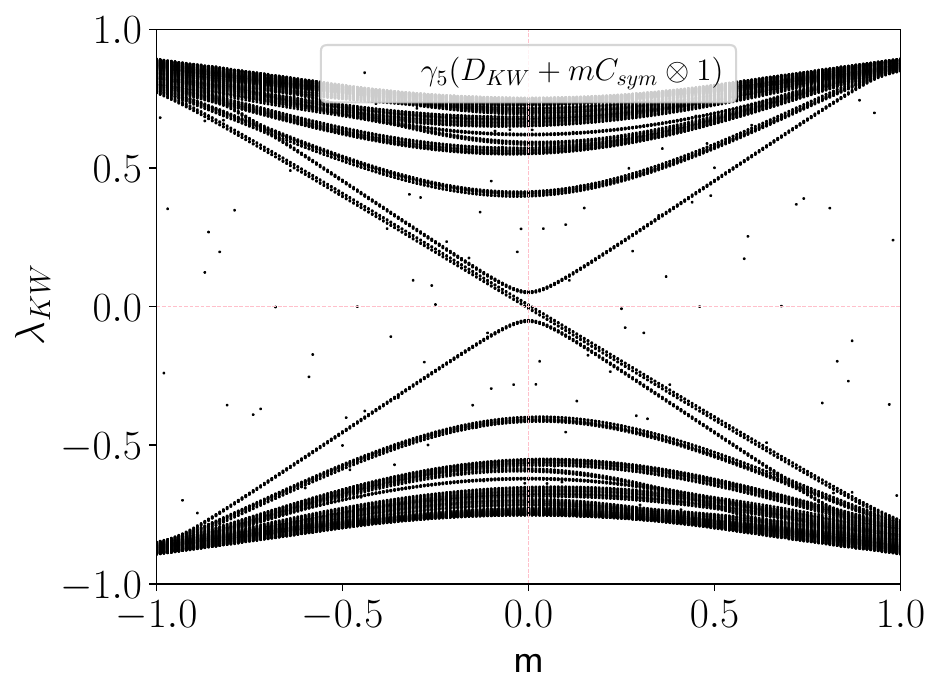} \hspace{0.1in}
\includegraphics[width=0.45\textwidth]{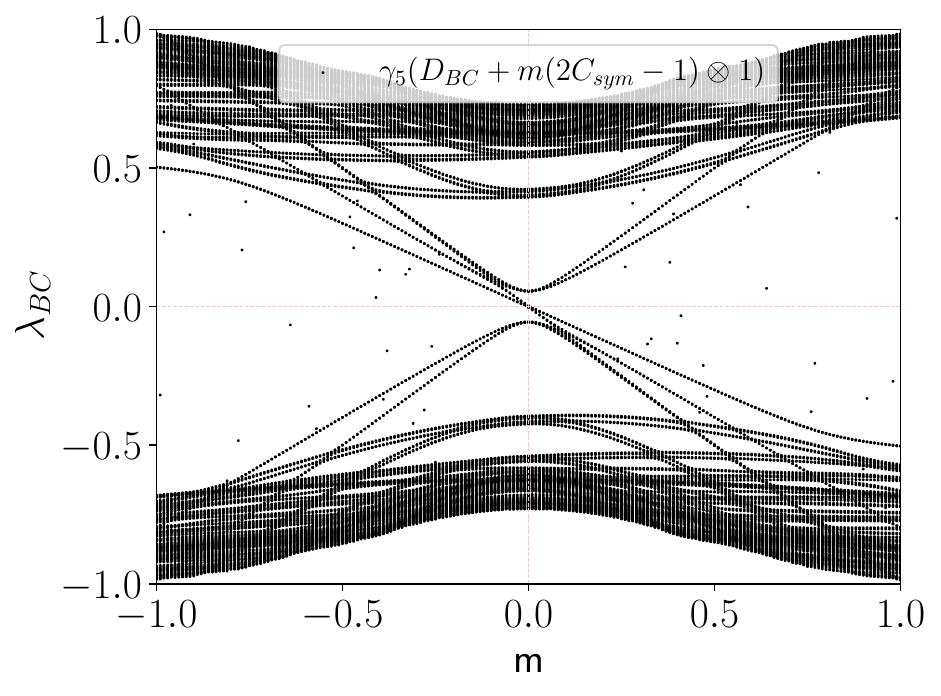}}
\vspace{-0.2in}
\caption{Spectral flow with respect to flavored masses of KW and BC fermion
on $8^4$ lattice with $Q_{\text{top}} =-2$ and $\delta=0.05$. Eigenvalues
in left and right panels correspond to $H_{\text{KW}}$ (eqn. \ref{hkw5})
and $H_{\text{BC}}$ (eqn. \ref{hbc5}), respectively.}
\label{fig:specflow}
\end{figure}
In Fig. \ref{fig:specflow} the flow of eigenvalues of $H_{\text{KW}}$
and $H_{\text{BC}}$ fermions with flavored mass are shown. When
eigenvalues are computed with bare quark mass instead of flavored mass,
\textit{i.e.,} with $\gamma_5(D_{\text{mdf}}+m)$, it shows the net crossing to be zero
as the would-be zero modes of either KW or BC Dirac operator cancel
between species doublers. This gets resolved with the use of flavored mass,
which is similar to what has been observed with staggered fermions~\cite{Adams:2009eb,Adams:2010gx}. We see two doubled crossings for
each of BC and KW around $m=0$ with negative slopes corresponding to
zero modes implying $\text{index}(D_{\text{mdf}})=2 Q_{\text{top}}=-4$.
The $Q_{\text{top}}$ calculated from the eqn. (\ref{qtop_def})
using the gauge field(s) generated is $Q_{\text{calc}} = -1.742$.

\section{Chiralities}\label{sec:chiralities}
In this section, we present the details of the eigenvalues and eigenvectors
along with measured $\gamma_5$ and modified $\gamma_5$-chiralities $X$,
\begin{equation}
X_{\text{KW}}=[C_{\text{sym}}\otimes\gamma_5] \;\;\;\text{and} \;\;\;
X_{\text{BC}}=[(2C_{\text{sym}}-1)\otimes\gamma_5] \label{xchi}
\end{equation}
Following \cite{Blum:2001qg}, in the first step, we
determined $N_{\text{max}}$ eigenvectors $\vert \psi_k \rangle$,
corresponding to that many lowest eigenvalues of $D_{\text{mdf}}^2$. For this, we used the Kalkreuter-Simma algorithm.
In the next step, a reduced $N_{\text{max}} \times N_{\text{max}}$ MDF 
Dirac operator is created by $\langle \psi_j\vert D_{\text{mdf}}
\vert \psi_k \rangle$. Subsequently, LAPACK is used on the reduced
$D_{\text{mdf}}$ matrix to obtain our desired
eigenvalues $\lambda_{\text{mdf}}$. The same eigenvectors are 
used to obtain $\langle \psi_i, \gamma_5 \psi_i \rangle$
and $\langle \psi_i, X_{\text{mdf}} \psi_i \rangle$.

\begin{table}[h]
	\centering
	\begin{tabular}{ | c || c  c  c || c  c  c | } \hline
		S.No & $\lambda_{KW}$ &
		$\langle \psi_i,\gamma_5\psi_i \rangle$ &
		$\langle \psi_i,X_{KW}\psi_i \rangle$   &
		$\lambda_{BC}$ &
		$\langle \psi_i,\gamma_5\psi_i \rangle$ &
		$\langle \psi_i,X_{BC}\psi_i \rangle$\\ \hline
		1 & 0.004962i &  0.00 & -0.80 & 0.000636i & 0.00 & -0.80\\
		2 & -0.004962i & 0.00 & -0.80 & -0.000636i & 0.00 & -0.80\\
		3 & 0.006934i & 0.00 & -0.80 & 0.004267i & 0.00 & -0.78\\
		4 & -0.006934i & 0.00 & -0.80 & -0.004267i & 0.00 & -0.78\\
		5 & 0.049193i & 0.00 & 0.05 & 0.054355i & 0.00 & 0.06\\
		6 & -0.049193i & 0.00 & 0.05 & -0.054355i & 0.00 & 0.06\\
		7 & 0.049675i & 0.00 & 0.05 & 0.055118i & 0.00 & 0.06\\ \hline
	\end{tabular}
	\caption{ Eigenvalues and computed chiralities of massless
		$D_{\text{KW}}$ and $D_{\text{BC}}$ for $Q_{\text{top}}=-2$ and $\delta=0.05$
		background $8^4$ lattice.} \label{table:chirality} 
\end{table}

\begin{figure}[H]
	\centering
	\makebox[\textwidth]{
		\includegraphics[width=0.40\textwidth]{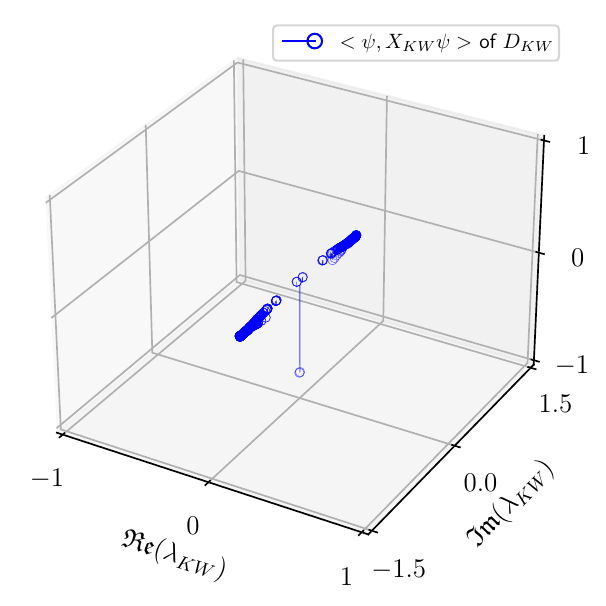} \hspace{0.1in}
		\includegraphics[width=0.40\textwidth]{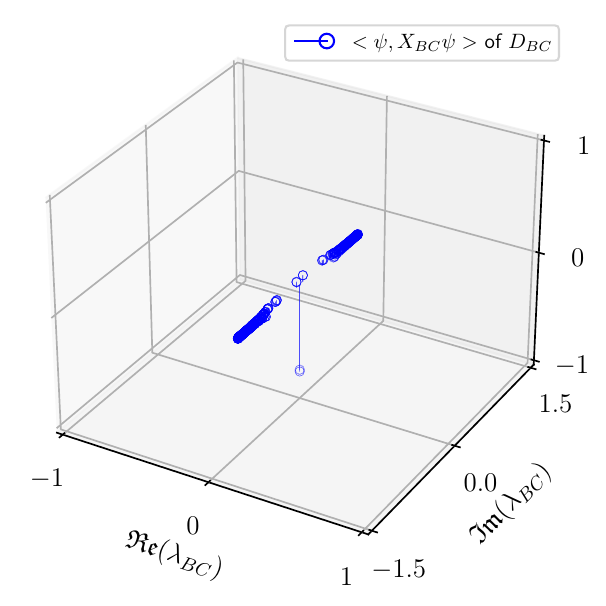}}
	\vspace{-0.2in}
	\caption{Needle plots of the modified $\gamma_5-$chiralities for massless $D_{\text{KW}}$ and $D_{\text{BC}}$ for $Q_{\text{top}}=-2$ and $\delta=0.05$ background $8^4$ lattice.
    The $\lambda_{\text{mdf}}$
	are concentrated on the imaginary axis as tabulated in the Table.
	\ref{table:chirality}, where the vertical axis corresponds to $\langle
	X_{\text{mdf}} \rangle$ with the stand-alone points reaching out to $-1$.}
	\label{fig:needle_plot}
	\vspace{-0.2in}
\end{figure}

In Table \ref{table:chirality}, we tabulated the first few lowest eigenvalues and their corresponding chiralities.
We found the zero eigenvalues $\lambda_{\text{mdf}}$ are of
$\mathcal{O}(10^{-3})$, whereas those of the first excited states are
$\mathcal{O}(10^{-2})$. We also found that $\langle \gamma_5 \rangle$
is always zero regardless of whether it is a zero eigenmode or not,
implying that $\gamma_5$ is not an appropriate measure to capture
the $\pm 1$ chirality corresponding to the zero eigenvalues. On the
other hand, the modified $\gamma_5$-chirality operator  $X_{\text{mdf}}$
(\ref{xchi}) correctly reproduced the chirality -0.80 ($\approx -1$) for four zero eigenmodes, implying
$\text{index}(D_{\text{mdf}})=-4$ which equals $2Q_{\text{top}}$, validating the index theorem.

The corresponding needle plots for KW and BC for modified chirality operators are
given in Fig. \ref{fig:needle_plot}, which, as is expected from the table,
show $\langle X_{\text{mdf}} \rangle \rightarrow \approx -1$ for four zero eigenmodes and $\langle X_{\text{mdf}} \rangle \rightarrow \approx 0$ for non-zero eigenmodes. The plots
are almost identical to those obtained in \cite{Durr:2022mnz}, showing no
new physics for KW or BC emerging in 4-dim.

\section{Eigenspectra with flavored mass}\label{sec:complex_eigenvalues}
We have seen above the operator $X_{\text{mdf}}$ can identify the
chiralities of the zero modes of the eigenstates of the massless KW and
BC Dirac operators in 4-dim. It now remains to see how the
tastes of the operators $(D_{\text{KW}} + mC_{\text{sym}}\otimes1)$ and $(D_{\text{BC}} + m(2C_{\text{sym}}-1)\otimes1)$ are
separated based on their chiralities. A previous study in 2-dim
\cite{Durr:2022mnz} showed that a flavored mass term successfully removes
the degeneracy in tastes. Here, too, we generated reduced matrices for
$(D_{\text{KW}} + mC_{\text{sym}}\otimes1)$ and $(D_{\text{BC}} + m(2C_{\text{sym}}-1)\otimes1)$ for $m=1$ using corresponding eigenvectors $\vert \psi_k \rangle$ of $(D_{\text{KW}} + mC_{\text{sym}}\otimes1)^2$ and $(D_{\text{BC}} + m(2C_{\text{sym}}-1)\otimes1)^2$, respectively, and determined
the eigenvalues of these reduced matrices.

\begin{figure}[htp]
\centering
\makebox[\textwidth]{
\includegraphics[width=0.45\textwidth]{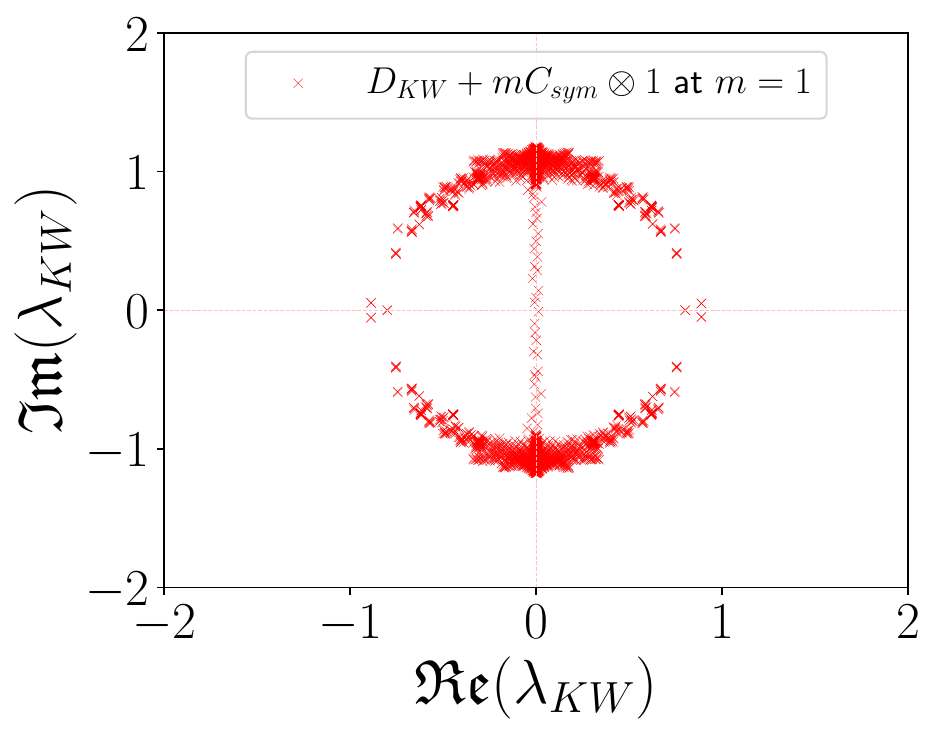} \hspace{0.1in}
\includegraphics[width=0.45\textwidth]{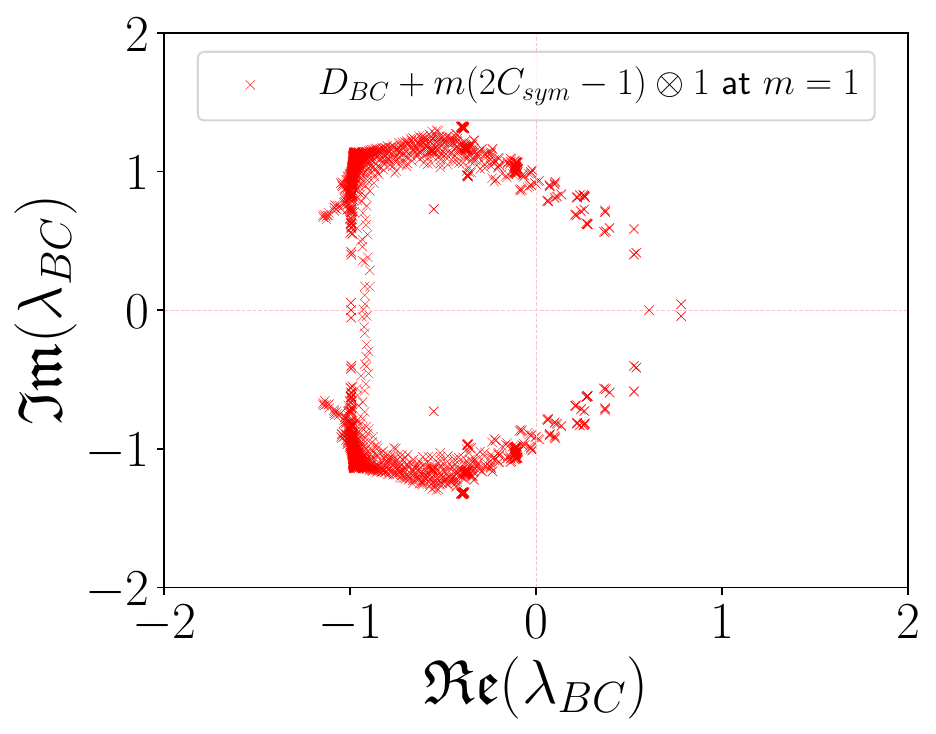}}
\vspace{-0.2in}
\caption{Complex eigenvalues of KW and BC Dirac operators with flavored mass terms: the eigenspectra split in two branches crossing the real axis at $m=\pm1$.}
\label{fig:cmpklx_eval}
\end{figure}
Fig. \ref{fig:cmpklx_eval}, shows that the
modified mass term separated the doublers (or tastes) according to their
$\pm1$ chirality appearing at $|m|$ on the real axis. However, for KW fermion, some
concentrations of eigenvalues are still at $m=0$, which possibly
result from using a small lattice volume and are expected to vanish for larger lattices.

\section{Conclusions}
The above set of observations indicates that the eigenspectra of the two
variants of minimally doubled fermions, namely Karsten-Wilczek and
Borici-Creutz, in 4-dim spacetime, is consistent with the index
theorem. For this, we generated gauge field configuration, subjected to
appropriate roughening, with a fixed topological charge $Q_{\text{top}}=
-2$. The charge is cross-checked by recalculating it using eqn. (\ref{qtop_def}).
Using flavored mass terms, we obtained the separation of doubler tastes of both KW and BC according to their chirality.

\section*{Acknowledgments}
\noindent The numerical jobs have been run on the HPC and Param
Sanganak at IIT Kanpur funded by DST and IIT Kanpur, and on the
computer facility at NISER, Bhubaneswar. A.K. would like to thank
Radhamadhab, Indrajit, Diptarko, Manisha and Abhishek. He would
further like to thank Stephan D{\"u}rr for useful discussions and
Taro Kimura for useful communications.

\bibliographystyle{JHEP}
\bibliography{ref}

\end{document}